# Navigating in Virtual Reality using Thought: The Development and Assessment of a Motor Imagery based Brain-Computer Interface

Behnam Reyhani-Masoleh[1, 2] and Tom Chau[1, 2, *]

[1] Bloorview Research Institute, Holland Bloorview Kids Rehabilitation Hospital, Toronto, ON, Canada, [2] Institute of Biomaterials and Biomedical Engineering, University of Toronto, Toronto, ON, Canada

* Corresponding author. Bloorview Research Institute, Holland Bloorview Kids Rehabilitation Hospital, 150 Kilgour Road, Toronto, Ontario M4G 1R8, Canada

*E-mail address:* tchau@hollandbloorview.ca

## Abstract

Brain-computer interface (BCI) systems have potential as assistive technologies for individuals with severe motor impairments. Nevertheless, individuals must first participate in many training sessions to obtain adequate data for optimizing the classification algorithm and subsequently acquiring brain-based control. Such traditional training paradigms have been dubbed unengaging and unmotivating for users. In recent years, it has been shown that the synergy of virtual reality (VR) and a BCI can lead to increased user engagement. This study created a 3-class BCI with a rather elaborate EEG signal processing pipeline that heavily utilizes machine learning. The BCI initially presented sham feedback but was eventually driven by EEG associated with motor imagery. The BCI tasks consisted of motor imagery of the feet and left and right hands, which were used to navigate a single-path maze in VR. Ten of the eleven recruited participants achieved online performance superior to chance ($p < 0.01$), while the majority successfully completed more than 70% of the prescribed navigational tasks. These results indicate that the proposed paradigm warrants further consideration as neurofeedback BCI training tool. A paradigm that allows users, from their perspective, control from the outset without the need for prior data collection sessions.

# 1. Introduction

Individuals with severe motor impairments often experience limitations in their ability to interact with their environment [1]. Severe motor impairments may be caused by various neurological, neuromuscular or neurodegenerative conditions [2]. In recent years, brain-computer interface (BCI) systems have shown potential as a means of re-routing commands originating from the brain to external devices, bypassing the spinal cord, peripheral nervous system and musculoskeletal system. By decoding an individual's intentions from recorded brain signals, BCI systems can provide hands-free control of mobility aids [3, 4].

A number of different mental tasks have been employed with BCIs, each one eliciting distinct and discernible patterns in recorded brain signals. Of these tasks, motor imagery (MI) – imagined movement of a specific body part – produces brain activity similar to that observed during motor execution [5]. Due to this natural relation, MI is a popular mental task used by BCIs for control applications (e.g., controlling a motorized wheelchair) [6]. However, to acquire adequate control of a MI-based BCI, individuals must first participate in long training procedures, which can persist for months [7]. The objective of these training procedures is to improve the user's ability to consciously elicit certain neurophysiological patterns by performing MI tasks. At the same time, the brain's electrical activity is collected and analyzed for patterns associated with the different MI tasks [8]. Unfortunately, even with these extensive training procedures, BCI performance is subpar compared to the level needed for use in the real-world [9], a setting where BCI error can lead to catastrophic events. This has led many to believe that the traditional training paradigms used by MI-based BCIs are suboptimal and perhaps even counter-productive [10]. Specifically, it has been shown that users are often bored, unengaged, unmotivated and frustrated during the early stages of training [11]. Being in these negative psychological states has shown to impede progress as it leads to impaired MI performance and the recording of irrelevant data. This cacophonous data obscures the underlying patterns associated with MI [12, 13]. To circumvent this issue, research groups have made use of machine learning (ML) techniques to successfully decode MI states from the noise-ridden data [14]. Furthermore, many BCIs have adopted a game-based approach to training to improve user experience. Some have even used virtual reality (VR) and have reported increased user engagement and motivation (refer to [15] for a review).

It has been suggested that the training protocol is the most crucial component in the design of a successful MI-based BCI [16]. For this reason, we proposed a paradigm that has a game-based VR application at its core, which provides feedback to the user prior to the development of a personalized brain signal classifier. VR is thus exploited to maintain the engagement and motivation of users throughout training [17] while ML is leveraged to uncover robust patterns associated with MI from the limited, recorded brain activity data. In summary, in this work, we developed and assessed a novel training paradigm for MI-based BCIs, which is engaging from the outset (i.e., before the machine is trained), by leveraging the complementary benefits of VR and ML.

## 2. Methods

### 2.1. Participants

Eleven able-bodied, naïve BCI users (9 males, 9 right-handed) between the ages of 20 and 30 (mean age: 24.7 ± 2.7 years) took part in the study after providing informed written consent. All participants were fluent in English, had normal or corrected-to-normal vision, and were asked to refrain from consuming caffeinated or alcoholic beverages for at least 4 hours prior to a session. Participants were free of degenerative, cardiovascular, psychiatric, neurological, respiratory, drug or alcohol-related conditions. In addition, participants were screened for history of epilepsy, seizures, photosensitivity, traumatic brain injury or concussions to mitigate the risk of VR-induced seizures [18]. Furthermore, to comply with the safety recommendations made by the VR headset manufacturer, participants confirmed that they were free from emotional stress, anxiety, digestive issues, pacemakers and hearing aids at the time of participation. Experimental protocols were approved by the ethics committee of the Holland Bloorview Kids Rehabilitation Hospital and the University of Toronto.

### 2.2. Brain-Computer Interface and Virtual Reality System Overview

Electrical signals representing brain activity were acquired using 16 active, dry electroencephalography (EEG) electrodes densely placed around the Sensorimotor cortex (SMC) region and amplified using a BrainAmp DC amplifier (Brain Products GmbH, Germany). The right and left mastoids were respectively used as the reference and ground electrodes. The electrode configuration, shown in Figure 1, was judiciously chosen to maximize the number of relevant electrodes placed near the SMC region while still allowing for comfortable use of the VR headset. Specifically, electrodes Fz, FC3, FCz, FC4, C5, C3, C1, Cz, C2, C4, C6, CP3, CP1, CPz, CP2, and CP4 were used.

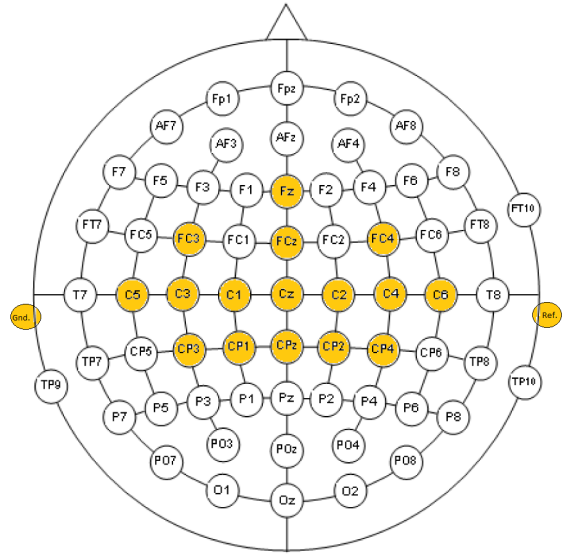

Figure 1. EEG electrode configuration

The Oculus Rift VR headset (Facebook Inc., California) was used because of its immersive experience and state-of-the-art technology. It requires computationally expensive graphics processing to render an immersive experience. Thus, relying on one computer for both VR graphics processing and EEG data processing could compromise the performance of both systems, thereby diminishing the participant's BCI experience. Therefore, two computers were deployed in the BCI-VR system: one for VR graphics processing and one for EEG data analysis. The GeForce GTX 1060 (Nvidia Corp., California) graphics processing unit was used as recommended by the Oculus Rift's specifications. Information was communicated from the VR computer to the EEG computer through the parallel port. Conversely, information was sent from the EEG computer to the VR computer through a network connection using the User Datagram Protocol (UDP). The BCI-VR system's components and their relation to one another is depicted in Figure 2.

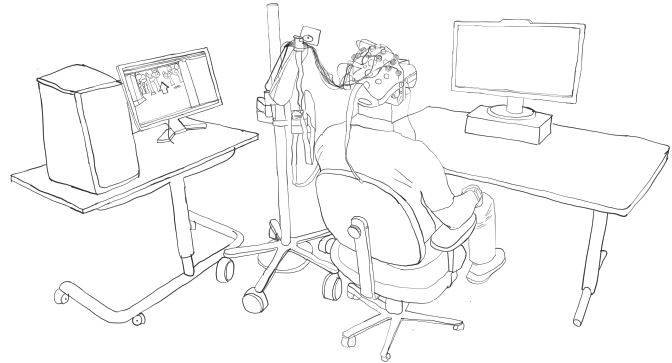

Figure 2. BCI-VR System Overview

## 2.3. Virtual Reality Application

The VR application was developed using the Unity game development platform (Unity Tech., California) with scripts written in C#. The setting of the application was a unidirectional maze situated in a forest. A forest setting was chosen to decrease stress and increase positive affect [19]. A birds-eye view of the maze and a first-person perspective of the application are shown in Figure 3.

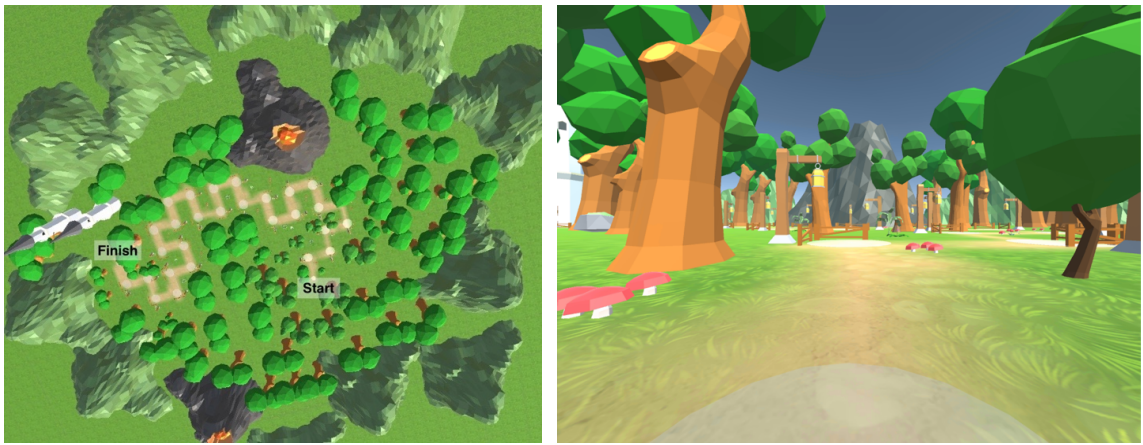

Figure 3. VR environment from a birds-eye view (left) and first-person perspective (right).

The path through the maze was punctuated with 29 checkpoints. Adjacent checkpoints (denoted as white circles in the left panel of Figure 3) were situated 5 meters apart in the virtual environment. At the first checkpoint, the user simply had to advance forward to the second checkpoint. At the remaining 28 checkpoints, the user could view the next checkpoint by rotating the VR character's field of view 90 degrees to either the left or right. Thus, to reach the subsequent checkpoint, the character had to first turn 90 degrees in one direction (14 checkpoints required turning to the left and 14 to the right) and then move forward 5 meters.

It is generally best practice in VR application development to avoid dynamic and continuous changes in perspective (i.e., the character's field of view moves or rotates continuously) as this can induce motion sickness [18]. As a remedy, many VR applications make use of teleportation [20]; this is the approach we took in our VR application. More specifically, to move forward to the next checkpoint, the teleporter – a multi-colored circular VR user interface element as shown in Figure 4 – had to move forward until it was situated within the ensuing checkpoint. The VR character was then teleported. The teleporter element's velocity was set to 5 meters/6 seconds and as a result, in the best possible scenario it would take 6 seconds for one to advance to the next checkpoint. We found that 6 seconds was an ideal minimum time for a task trial; this duration was sufficient to challenge the participant without risk of boredom. The position of the teleporter element at different times of a "move forward" task is shown in Figure 4.

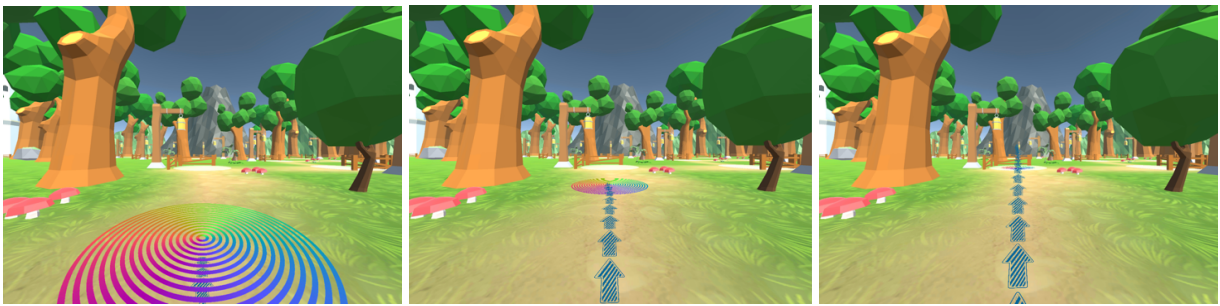

Figure 4. Move forward task; VR teleporter element at various positions during the trial: 0 meters, 3 meters, and 5 meters

To rotate one's field of view 90 degrees to the left or right, a similar approach was taken. In these scenarios, the VR arrow element would appear, as shown in Figure 5. To rotate one's field of view, the arrow element would have to pivot 90 degrees in the intended direction. The arrow's rotational speed was set to 90 degrees/6 seconds, and therefore in the best case, one's field of view would change after 6 seconds. Figure 5 depicts the VR character's viewpoint at various times within a "rotate right" task.

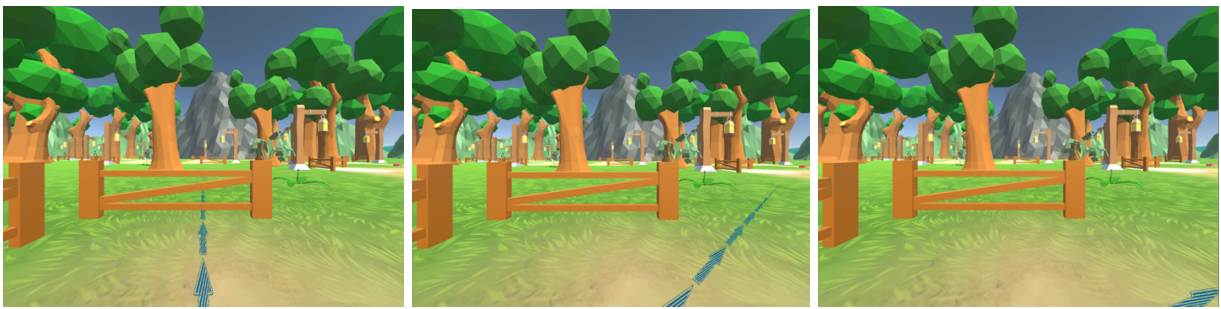

Figure 5. Rotate right task; VR arrow element at various angles during the trial:
0 degrees, 45 degrees, and 85 degrees

## 2.4. Experimental Protocol

During each phase of the experimental protocol, participants sat comfortably in a chair with the EEG cap and VR headset securely fitted on their head. Participants attended two sessions in total on separate days.

### 2.4.1. Pre-Session Practice

The purpose of the pre-session phase was three-fold: 1) have the user practice kinesthetic-MI, 2) become comfortable with the BCI-VR set up, and 3) acclimatize to being immersed in the virtual environment. Participants were first introduced to the concept of kinesthetic-MI and its difference from visual-MI [89]. Specifically, it was emphasized that kinesthetic-MI involves the experience of the bodily sensations associated with the movement from the first-person perspective. Afterwards, while wearing the EEG cap, participants were placed inside the virtual environment and instructed to perform movement of the right hand, left hand or both feet. Each participant performed 15, 6 second trials for each of the three MI tasks.

### 2.4.2. Offline Calibration Session – Collecting Training Data

In the offline calibration session, the participants' brain activity was recorded while they performed the three MI tasks. Specifically, they were told that by performing MI tasks they could control the VR application: feet MI moves the VR teleporter element forward; and right (left) hand MI rotates the VR arrow element to the right (left). To increase their motivation, it was expressed to them that the goal was to complete the maze in the quickest time possible. Nevertheless, participants were presented with "sham" feedback: the VR movement elements, teleporter and arrow, would move or rotate irrespective of the participant's brain signals. To ensure that they had no reservations to whether the feedback was real, they were told that the data from pre-session practice was used to train an algorithm that could distinguish between the different MI tasks from the EEG recordings. Furthermore, if the algorithm detected a MI task that was congruent with the current task, then the VR element would move appropriately, otherwise the element would remain stationary. Additionally, the participants were told that the system was not perfect and therefore it was important that they continue to perform the correct MI task as they go through the maze, irrespective of what was displayed in the application. It should be emphasized that at the end of the study, it was revealed to the participants that they had received "sham" feedback in this session.

Each participant completed two runs through the maze. To determine the type of sham feedback to display at each instance, the VR application would draw from a Bernoulli distribution every 2 seconds: $P(n) = p^n(1-p)^{1-n}$ . If the sample drawn was a '1' then the VR movement element for the current task, i.e., VR teleporter or arrow, would move or rotate at a constant rate for the ensuing 2 seconds. This scenario represented a "correct classification" by the BCI. Conversely, if a '0' was drawn then the VR movement element would remain stationary; this represented a "misclassification" by the BCI. The probability hyperparameter, '$p$', was set to 0.65 for the first run through the maze and 0.75 for the second run. This paradigm was used to emulate an improving yet imperfect BCI. Each task would be successfully completed after three correct classifications. At the end of each maze run, participants were instructed to fill out a post-session survey.

### 2.4.3. Online – Real-Time Feedback

The collected EEG data from the offline calibration session was used to develop the BCI's EEG signal processing pipeline so that it could decode in real-time the MI task the participant was performing. The participants who appeared to have the potential to control their BCI at above chance level were invited back to an online session. These participants attained an average kappa score above zero during cross-validation. In this session, users were instructed to go through the maze as they had done before, using MI tasks as control commands for the VR application. Participants were told that a different classification algorithm was being used compared to the previous two runs and therefore it may be more difficult to control the BCI. This was done to mitigate the risk of them discovering that "sham" feedback was presented to them in the previous session. Discovery of the use of "sham" feedback may negatively affect their psychological state, perhaps leading to reduced motivation and performance during online operation. Each participant had 14 seconds to accomplish a task trial before the VR application proceeded to the next task of the maze. This maximum time for a trial was chosen such that the participant would spend at worst case 13.3 minutes in the maze (57 task trials, 14 seconds per task trial). Once again, participants were instructed to complete a post-session survey after the maze run.

## 2.5. EEG Signal Processing Pipeline

### 2.5.1. Pipeline Development using Calibration Data

The BCI's signal processing pipeline was developed on the basis of the EEG data collected from the offline calibration session. In turn, the pipeline was used in the online session to classify EEG signals in real-time.

### 2.5.2. Signal Preprocessing using Independent Component Analysis

The collected EEG data was sampled at 250 Hz and band pass filtered between 1 Hz and 45 Hz. Due to volume conduction effects [21], the recording at a single EEG electrode on the scalp is the aggregation of electrical activity from different neural processes and even non-neural processes (e.g., electrooculogram). As a result, the recorded EEG has poor spatial resolution and is artifact-ridden. Independent Component Analysis (ICA) has been widely applied to enhance spatial resolution and reduce artifacts [22]. We applied ICA as a form of signal preprocessing to the EEG signals using the infomax algorithm [23].

### 2.5.3. Feature Extraction using Periodogram

Power spectral density (PSD) estimation techniques have been comparable, at times even superior to other methods for extracting the distinctive neuro-physiological markers of different MI tasks [24]. The most economical PSD estimation method is the periodogram, which is primarily the Discrete-time Fourier Transform (DTFT) of a sequence of data points, $x(n)$, which has length $N$ and sampling frequency $F_{Samp}$:

$$P(f) = \frac{1}{N}|DTFT(x(n))|^2 = \frac{1}{N}\left|\sum_{n=0}^{N-1} x(n)e^{\frac{-j2\pi fn}{F_{Samp}}}\right|^2$$

It is very attractive for online operation, especially in paradigms which require very frequent and short computation. Due to the nature of our protocol, we could not afford using relatively long data sequences since participants spent at minimum 6 seconds in each trial. For this reason, we decided to use 2 seconds worth of data for PSD estimation. Specifically, we epoched the offline calibration data in 2 second time-blocks corresponding to the time between consecutive "sham" feedback decision points. Thus, each 2-second signal segment consisted of 500 data points for each of the 16 EEG electrodes. Each 2-second segment was then linearly transformed using ICA. For every signal segment and each of its ICA components, the PSD was estimated, zero-padding to a sample length of 512. Using these settings, PSD estimation yielded a frequency resolution of approximately 0.5 Hz. To reduce the number of features, adjacent bins were averaged. Lastly, PSD estimates in frequency bins between 2 Hz and 40 Hz were extracted for each ICA component and concatenated into a final feature vector.

### 2.5.4. Feature Selection using Mutual Information Scores

The extracted feature vector was high dimensional, and since the amount of training data was limited, the number of features were reduced to mitigate the risk of overfitting. Most of the PSD features were irrelevant to the classification task as they represented frequency bands across scalp locations that were not indicative of the general pattern associated with MI: focal event-related desynchronization and surround synchronization [25].

Feature selection filters are computationally fast and avoid overfitting compared to wrapper methods (e.g., genetic algorithm, sequential selection, etc.) [26, 27]. Therefore, we adopted the use of an initial feature selection filter to drop a number of superfluous features. Since this approach selects suboptimal sets that comprise of redundant features, we decided to also incorporate an embedded feature selector in the classification stage of our pipeline; a two-step approach that has been recommended for datasets that have more dimensions than data samples [26]. An efficient and simple feature selection filter is one based off of the mutual information score between the target label, '$y$' and a feature '$x_i$', which is defined as:

$$Mutual\ Information\ (y, x_i) = I[y, x_i] = H[y] - H[y|x_i]$$

where $H[\bullet]$ is the information entropy. Information entropy quantifies the average amount of information received by observing a realization of a random variable. In summary, mutual information was estimated between each feature and the target label to quantify the amount of information received about the target label through observing the value of that specific feature. The $k$ features with the highest mutual information were selected.

### 2.5.5. Classification and Embedded Feature Selection using Linear Support Vector Machine with L1 Penalty

Linear Support Vector Machines (SVM) have been successful in MI task classification [28, 29]. They are effective in cases where the number of dimensions is greater than the number of samples and are less prone to overfitting [30]. To further reduce the number of effective features, we have adopted the use of the L1-norm penalty [31], which performs an embedded feature selection by driving feature weights to zero. This is not possible when using the standard L2-norm penalty [30]. For a binary classification problem, given $N$, training vectors $x_i$ and the corresponding class labels $y_i \in \{-1, +1\}$, the linear SVM with the L1-norm penalty looks to solve:

$$\min_{\mathrm{w,b,\zeta}} \sum_{i=1}^{N} C\zeta_i + \|\mathrm{w}\|_1$$

$$subject\ to: y_i(w^T x_i + b) \geq 1 - \zeta_i \quad and \quad \zeta_i \geq 0 \ , \quad i = 1, \dots , N.$$

Afterwards, new training examples can be classified using the decision function:

$$f(x) = sgn(w^T x + b)$$

To support multi-class classification, we used the "one-vs-all" scheme. In this paradigm, a classifier is trained for each class, where training examples belonging to that class are labelled the positive class while all remaining samples are pooled into the negative class. To account for our unbalanced dataset, we replaced the penalty parameter, $C$, in the above formula with:

$$C_{new} = C \times W_i$$

where $W_i$ is the fraction of training examples belonging to the $i^{th}$ class, and $\sum_i W_i = 1$ . We standardized each feature to 0 mean and unit variance.

### 2.5.6. Optimizing Pipeline Hyperparameters

The pipeline had two hyperparameters: the number of features, $k$, selected by the mutual information feature selection filter and the penalty parameter, $C$, of the linear SVM cost function. To find an estimate for the optimal hyper-parameters, we used a stratified 5-fold cross validation which was repeated 10 times.

### 2.5.7. Online Operation – Real-Time Motor Imagery Classification from EEG

With the pipeline developed and the classifier trained, the BCI was deployed for online operation. Every second, the pipeline was fed with 2 seconds of recorded EEG data. The data was processed by the pipeline to infer which one of the three MI tasks was being performed by the participant. This decision was then sent to the VR application through the network to modulate the feedback to the user. Figure 6 depicts the BCI pipeline and the data analysis paradigm used for online operation.

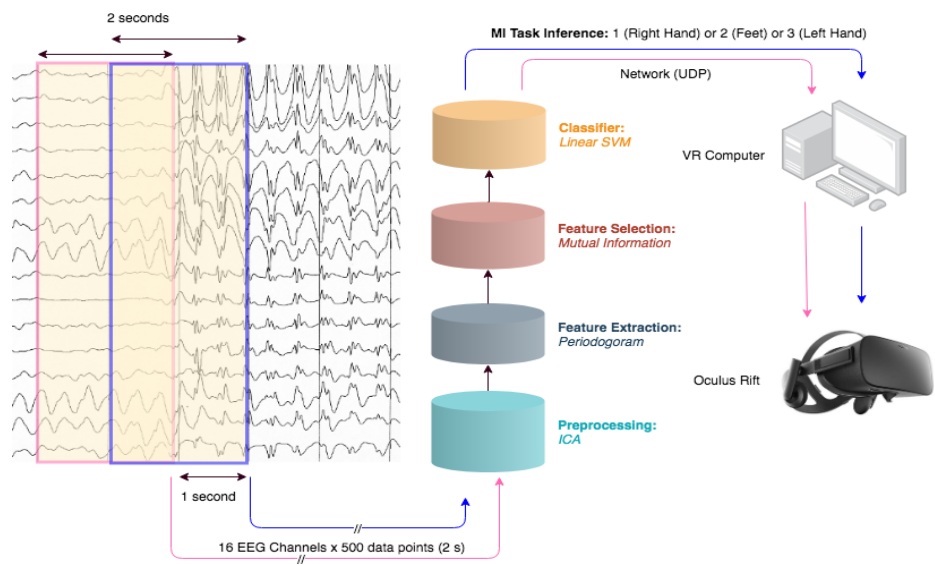

Figure 6. BCI pipeline and online operation paradigm

### 2.5.8. Monte Carlo Simulation to Obtain Performance Metric Benchmarks

To obtain benchmark performances for online operation, we simulated runs with two "dummy" classifiers, i.e., classifiers that do not take into account input data. This allowed us to juxtapose the online performance of participants with these classifiers. '*Dummy-Stratified*' generated random predictions by respecting the training set class distribution. Secondly, '*Dummy-Uniform*' generated predictions uniformly at random. We ran the two classifiers through 100,000 simulated runs to obtain means and standard deviations for the performance metrics under the respective null hypotheses of uniformly random and a priori probability-weighted random classification. Afterwards, the z-scores for the performance metrics of each participant were compared to the two random classifiers' chance distributions.

## 3. Results

All participants successfully completed over half of the task trials in the maze. Recall that the maze consisted of 57 total task trials. P8 was the worst performer, only successfully completing 52.63% of trials. On the other hand, P3 had the most successful trials, 89.47% to be exact. On average, participants completed 71.93% ±12.92% of the 57 trials. Only P5, P7 and P8 failed to finish more than 65% of the trials. Four out of the ten participants (P2, P3, P6, and P9) achieved 80% trial completion. All participants achieved superior task completion compared to *Dummy-Uniform* at *p < 0.01.* Similarly, all but P7 exceeded *Dummy-Uniform* task completion (*p < 0.05).* The percentage of successfully completed trials for each participant and the benchmarks are shown in Figure 7. Double asterisks or plus signs indicate statistically significant (*p < 0.01*) values compared to *Dummy-Stratified* and *Dummy-Uniform.* A single symbol indicates statistical significance at *p < 0.05*.

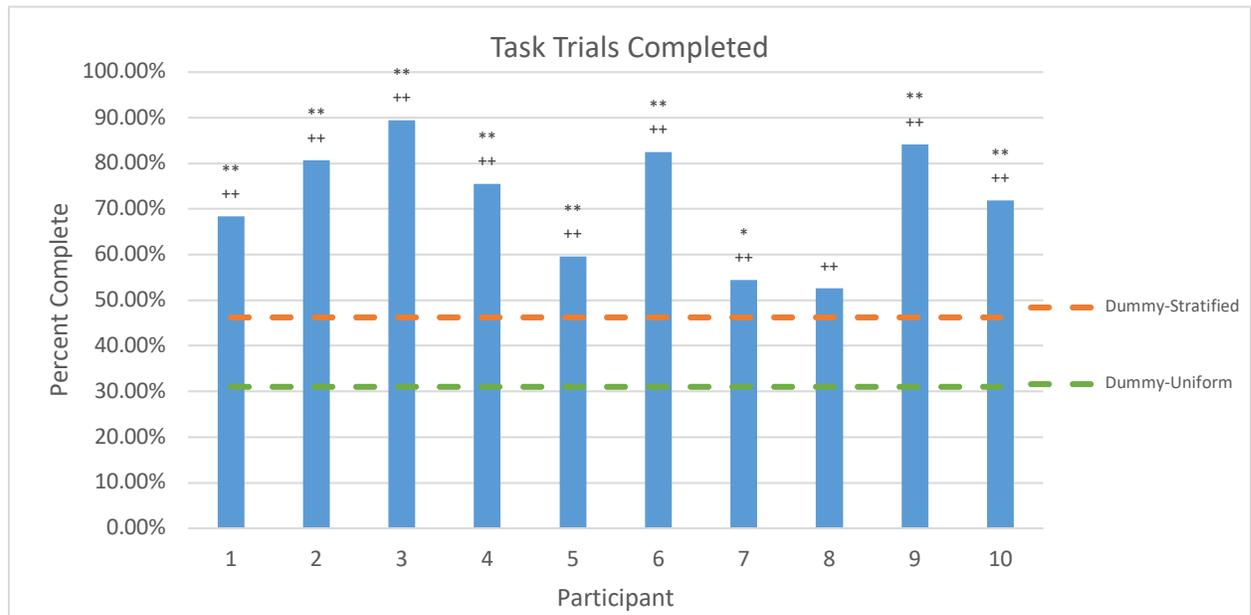

Figure 7. Percentage of successfully completed task trials: Double symbols '**' and '++' indicate significance difference at *p < 0.01*, while single symbols '*' and '+' denote significant difference at *p < 0.05*, when compared to stratified (*) and uniform (+) dummy classifiers.

The participants' performance varied across the different MI tasks. Of the 14 right-hand MI trials in the maze, the number of successfully finished trials varied from 6 to 14 across participants. On average, participants successfully finished $10 \pm 3$ right-hand MI trials in the maze. All participants successfully completed more right-hand MI task trials than either of the two random classifiers, i.e., *Dummy-Stratified* and *Dummy-Uniform*. Specifically, all participants achieved superior performance ($p < 0.01$) for right-hand MI compared to *Dummy-Stratified*. On the other hand, all participants except for P1 and P7, achieved superior ($p < 0.01$) right-hand MI performance compared to *Dummy-Uniform*. These results for task completion for right-hand MI are depicted in Figure 8.

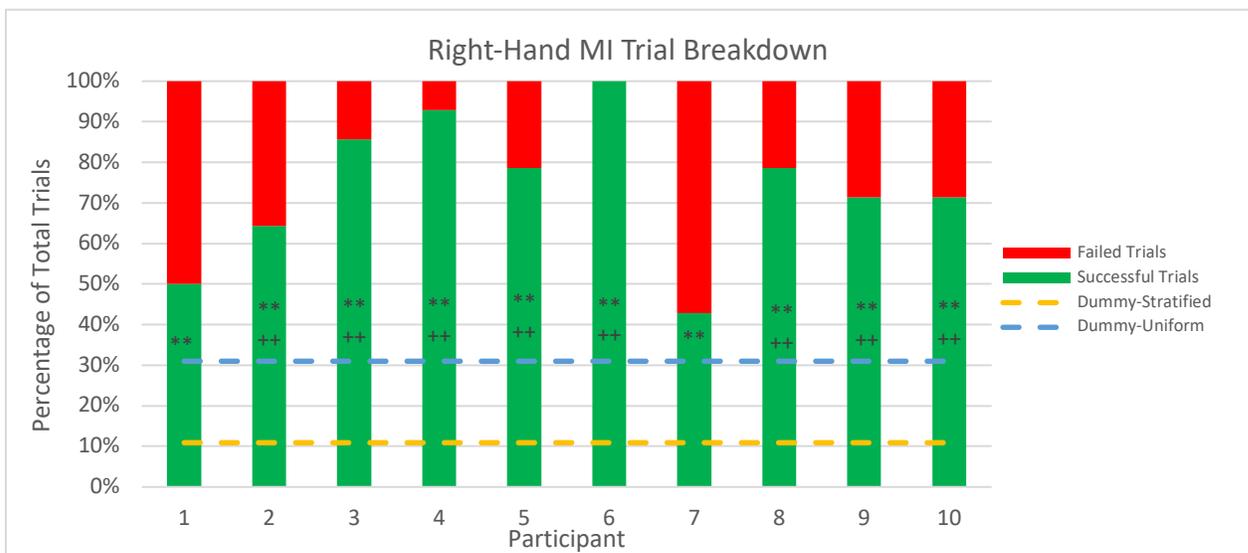

Figure 8. Fraction of Successful (Green) and Failed (Red) Trials for Right-Hand MI Tasks: '**'and '++' represent superior scores ($p < 0.01$) compared to those achieved by Stratified and Uniform classifiers

With respect to the feet MI task, the number of successful trials varied from 12 to 28 out of 29 trials, and on average $20 \pm 5$ trials were successfully completed. Only four participants outperformed the theoretical benchmark set by *Dummy-Stratified*. With that being said, for only P3 was this superior performance statistically significant ($p < 0.05$). On the other hand, all participants except P8 scored higher ($p < 0.05$) than *Dummy-Uniform* for this task. From these participants, only P5's superior performance in comparison to *Dummy-Uniform* was not statistically significant with $p < 0.01$. The breakdown of feet-MI trial task performance is plotted in Figure 9.

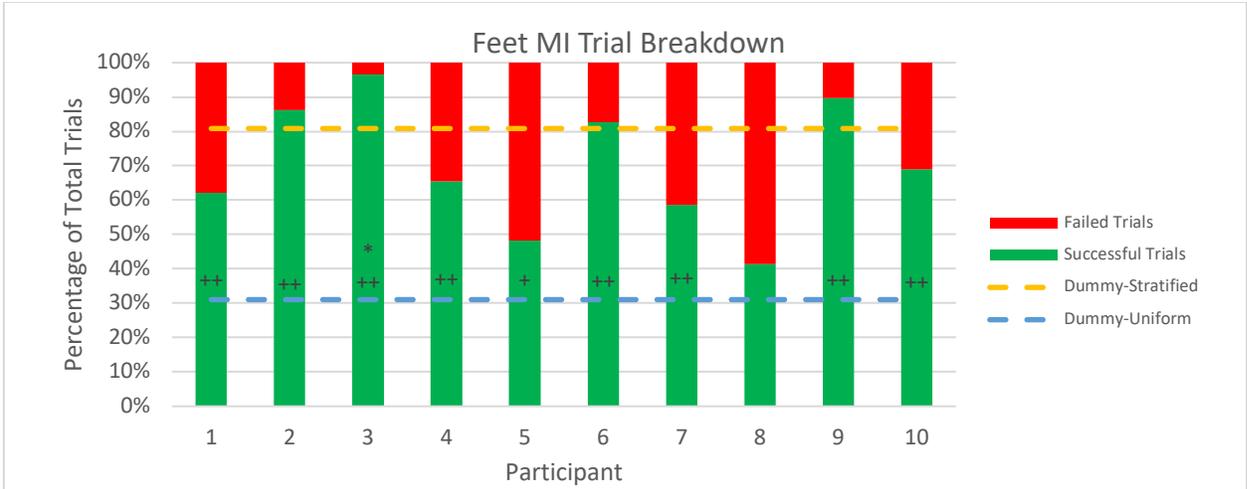

Figure 9. Fraction of Successful (Green) and Failed (Red) Trials for Feet MI Tasks: Double-markers '**'and '++' represent higher scores ($p < 0.01$) compared to Dummy-Stratified and Dummy-Uniform classifiers; likewise, single markers '*' and '+' indicate superior scores at $p < 0.05$.

Lastly, the number of successfully finished left-hand MI trials ranged from 7 to 14. On average, $10 \pm 2$ left-hand MI trials were successfully completed by participants. All participants scored better than the two theoretical benchmarks set by the random classifiers. Specifically, all participants surpassed the benchmark set by *Dummy-Stratified at p < 0.01*. On the other hand, only P8 did not achieve a better performance than *Dummy-Uniform*. These results are plotted in Figure 10.

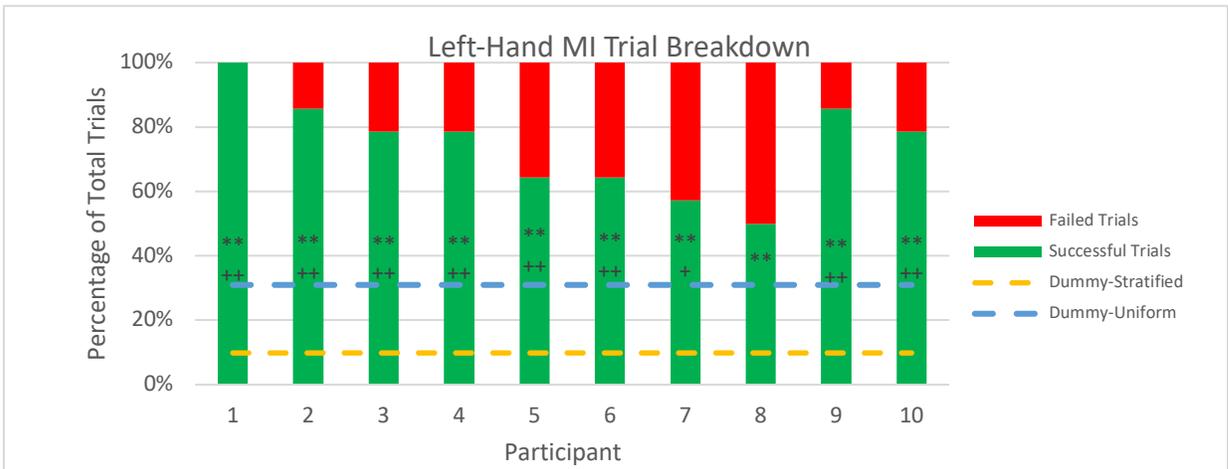

Figure 10. Fraction of Successful (Green) and Failed (Red) Trials for Left-Hand MI Tasks: '**'and '++' represent percentages that are greater ($p < 0.01$) than that attained by Dummy-Stratified and Uniform classifiers, and '*' and '+' indicate superior scores at $p < 0.05$.

# 4.  Discussion

The online performance achieved by participants in our study varied drastically; however, all participants exceeded chance-level benchmarks (at statistical significance with *p < 0.01*) obtained by Monte Carlo simulation using the random classifiers. Our results show that it is possible to obtain above chance-level online performance without any feedback training sessions. This is congruent with the Berlin-BCI's "let the machine learn" motto, where users have achieved high online performance using a MI-BCI without any prior feedback training sessions [14, 32, 33]. In addition, all participants were able to successfully complete over 50% of the prescribed tasks, with the majority completing over 70% of the tasks. Collectively, these findings suggest that VR-based interaction with initial sham feedback is a viable approach to promote MI training of naïve BCI users.

Inter-task performance variability was prevalent among participants. Specifically, some participants exhibited outstanding performance with one or two tasks, e.g., right-hand MI, but were unable to reliably perform the other tasks. P1 and P6 are extreme examples which best portray these phenomena. Specifically, P1 was able to complete all fourteen left-hand MI trials successfully but was only able to complete 60% of the feet-MI trials and 50% of the right-hand MI trials. On the other hand, P6 completed all right-hand MI trials successfully, but was only able to complete 80% of the feet-MI trials and 60% of left-hand MI trials. This is expected as the brain networks implicated in MI vary drastically across individuals, and the networks associated with MI of specific body parts vary considerably within an individual [34]. This is attributed to the activity-dependent plasticity of motor-related brain regions, which undergo dynamic reorganization with everyday experiences such as motor-skill learning and cognitive motor actions [35]. As a result, it is possible for an individual to elicit profound and discernible patterns for a specific MI task, e.g., right-hand MI, but not others; this phenomenon of preferential task-related MI activation has been identified as a distinguishing characteristic between expert athletes and non-athletes [36]. Interestingly, it has been shown that MI training also leads to similar effects, that is the re-organization of related brain regions [37-39]. Thus, to improve performance of a sub-par MI task (i.e., right-hand MI for P1 of our study), one should specifically target training of the task in question, with task-related feedback.

Comparisons across different BCI studies are not entirely valid as there are many differences between protocols. Our data collection session was drastically different from a traditional data collection session. Participants were immersed in a florid and visually appealing VR application, amid distractions such as continuously changing feedback elements. In contrast, traditional data collection sessions present no visual feedback to the users as they fixate on a visual anchor and barren or graphically primitive computer application while performing MI tasks. Indeed, the distractive nature of our application was reflected by the average distraction score, $3.81 \pm 0.86$, reported by the post-run surveys (Appendix A). Furthermore, a few of our participants reported that at times they became distracted as they would look ahead in the maze to determine which task was next. In relation, the literature supports the conjecture that a distracting paradigm can lead to reduced discernibility of MI patterns and hence, reduced performance. In a study that used a non-MI based BCI, it was shown that increased mental workload due to distractions reduced the BCI's control signal strength [40]. This distraction effect corroborates the finding that attention is a contributing factor to a BCI's performance; higher self-reported attention corresponded with higher performance [41]. It has also been shown that distractions significantly impact the pattern of MI-related brain activity. Specifically, the sustained MI activity is reduced amid distractions as users temporarily orient their attention to task-irrelevant stimuli before returning to the task at hand [42]. In the future, it may be better to decrease the length of the maze to sustain the user's attention and reduce the likelihood of distraction. This modification was suggested by a few participants, who claimed that the maze was lengthy. They recommended the use of multiple, different mazes that are shorter in length.

The use of shorter length mazes may also be beneficial for mitigating other negative psychological states such as mental fatigue. Specifically, it has been shown that increased mental fatigue as a result of prolonged time spent on a BCI task influences EEG PSD values, leading to reduced performance [43].  In our study, participants found the protocol somewhat fatiguing as observed by their survey results; the average mental fatigue score across 32 maze runs of the study was $3.72 \pm 0.99$. Interestingly, we found a strong negative relationship between a participant's average reported mental fatigue and their online performance ($r = -0.9125$). Indeed, fatigue is known to diminish BCI performance [13, 41]. Motivation has also been shown to affect a user's BCI performance; higher motivation has been correlated with improved BCI performance [12, 44]. Similarly, we found a strong positive linear relationship between our participants' average reported motivation and their online performance (*r = 0.8721*). In general, the grand average of the motivation score from the survey results, $3.63 \pm 1.18$, indicated that participants were motivated, which has been emphasized as a pre-requisite for MI-based BCI training [11].

Lastly, the design of the online EEG signal processing pipeline can also impact performance. Specifically, in BCI design one usually encounters a trade-off: shorter time-windows for analysis tends to lead to reduced performance but increased BCI response rate. For example, many of the PSD estimation methods used for feature extraction are asymptotically unbiased [45]. In other words, the PSD estimate approaches the actual value with longer time-windows. This property is exhibited in the method we employed, the periodogram. In addition, using longer time-windows also allows the use of methods that reduce the variance of PSD estimates through averaging, for example, Welch's method. As mentioned earlier, we decided to trade-off BCI performance for more responsiveness due to the nature of our paradigm and thus used a shorter time-window, i.e., 2 seconds. Indeed, an increase in BCI performance as a result of using longer time-windows for extracting PSD features has been observed in MI-BCI studies [46].

## 5. Conclusion

We have proposed a novel and engaging paradigm for EEG-driven MI-based BCIs based on a goal-oriented VR application. The use of "sham" VR feedback during the initial offline calibration session permitted the immediate collection of task-relevant EEG data, while motivating and engaging the participant. Adopting this paradigm allows naïve BCI users control from the first session (from their perspective) without the need for prior training or data collections sessions. This leads to a much more fulfilling user experience, which mitigates the possibility that users become discouraged by the technology. This approach starkly contrasts traditional training paradigms where no feedback is initially presented, resulting in low quality data for machine training and minimal participant engagement. The former translates into inaccurate on-line classifiers while the latter can potentially lead to users rejecting further use of the BCI [47]. We showed that all participants were able to control the VR application in online operation at above-chance level using the 3-class BCI. More importantly, participants were able to complete the majority of prescribed tasks. Such results are promising, given that all participants had no previous experience using BCI technology. Thus, the proposed paradigm can be used as the basis for further online training where feedback is EEG-driven. An analysis of BCI performance at the level of individual tasks, as presented here, can be used to effectively tailor a training regimen to specifically emphasize training of tasks that elicit less discernible activations.

To conclude, by adopting the VR feedback paradigm presented here, BCI researchers will be able to design and share "mazes" among each other in the hope of establishing a standard test for evaluating MI-based BCI control in real-world environments. This is an attractive proposition for two reasons. Firstly, BCI performance is seldom tested outside laboratory settings and in environments that resemble the real-world. Secondly, the testing medium and tasks tend to vary among researchers making it virtually impossible to accurately compare BCI performance across studies.

# Appendices

Appendix A – Post Session Survey

Participant #:                                    Date:

Run Number:

Please rank how you felt the session was to the best of your abilities (1: Very low, 3: Neutral, 5: Very high):

1. Mental Fatigue:

        1              2              3              4              5

2. Distraction:

        1              2              3              4              5

3. Motivation:

        1              2              3              4              5

General comments regarding the session: